\documentclass[reprint, superscriptaddress, secnumarabic, amssymb, nobibnotes, aps, prl]{revtex4-1}

\usepackage{graphicx}
\usepackage{epstopdf}
\usepackage[T1]{fontenc}
\usepackage[latin9]{inputenc}
\usepackage{amsbsy}
\usepackage{gensymb}
\usepackage[T1]{fontenc}
\usepackage[latin9]{inputenc}
\usepackage{amsmath}
\usepackage{amssymb}
\usepackage{bbm}
\usepackage{braket}
\usepackage{xcolor}
\allowdisplaybreaks
\usepackage{graphicx}
\usepackage[colorlinks=true]{hyperref}  
\hypersetup{
    bookmarks=true,         
    unicode=false,          
    pdftoolbar=true,        
    pdfmenubar=true,        
    pdffitwindow=false,     
    pdfstartview={FitH},    
    pdftitle={NbReSi},    
    pdfauthor={},     
    pdfsubject={},   
    pdfcreator={},   
    pdfproducer={}, 
    pdfkeywords={} {} {}, 
    pdfnewwindow=true,      
    colorlinks=true,       
    linkcolor=blue, 
    citecolor=blue,        
    filecolor=magenta,      
    urlcolor=blue           
} 
\usepackage[normalem]{ulem}

\newcommand{\figref}[1]{Fig.~\ref{#1}}

\begin{document}
\title{\textrm{ Superconductivity in noncentrosymmetric NbReSi investigated by muon spin rotation and relaxation}}
\author{Sajilesh~K.~P.}
\affiliation{Department of Physics, Indian Institute of Science Education and Research Bhopal, Bhopal, 462066, India}
\author{K.~Motala}
\affiliation{Department of Physics, Indian Institute of Science Education and Research Bhopal, Bhopal, 462066, India}
\author{P.~K.~Meena}
\affiliation{Department of Physics, Indian Institute of Science Education and Research Bhopal, Bhopal, 462066, India}
\author{A.~Kataria}
\affiliation{Department of Physics, Indian Institute of Science Education and Research Bhopal, Bhopal, 462066, India}
\author{C.~Patra}
\affiliation{Department of Physics, Indian Institute of Science Education and Research Bhopal, Bhopal, 462066, India}
\author{A.~D.~Hillier}
\affiliation{ISIS Facility, STFC Rutherford Appleton Laboratory, Harwell Science and Innovation Campus, Oxfordshire, OX11 0QX, UK}
\author{R.~P.~Singh}
\email[]{rpsingh@iiserb.ac.in}
\affiliation{Department of Physics, Indian Institute of Science Education and Research Bhopal, Bhopal, 462066, India}

\date{\today}
\begin{abstract}
\begin{flushleft}
\end{flushleft}
Noncentrosymmetric materials are promising paradigm to explore unconventional superconductivity. In particular, several Re containing noncentrosymmetric materials have attracted considerable attention due to a superconducting state with a broken time reversal symmetry. A comprehensive study on the superconducting ground state of NbReSi was investigated using magnetization, resistivity, and muon spin rotation/relaxation measurements. Zero field muon spectroscopy results showed the absence of any spontaneous magnetic field below the superconducting transition temperature, T$ _{c} $ = 6.29 K, indicating the preserved time-reversal symmetry. Transverse field muon spin rotation measurements confirms a s-wave nature of the sample with $\Delta(0)/k_{B}T_{c} $ = 1.726. This study urges further investigation on more noncentrosymmetric materials to elucidate the selective appearance of unconventional nature and unveil its dependence on antisymmetric spin-orbit coupling strength.

\end{abstract}
\maketitle
\section{INTRODUCTION}
Despite decade-long research on understanding the pairing mechanism in unconventional superconductors, the experimental evidence of exotic superconducting properties in various systems is still under debate among condensed matter physicists \cite{EBA,smid}. The field has observed a surge in research interest since the discovery of coexisting superconducting and anti-ferromagnetic phase in heavy-fermion noncentrosymmetric compound CePt$ _{3}$Si, along with the unusual superconducting nature \cite{Bauer2004}. Recent evidence of protected topological surface states in superconductors has once again renewed the interest in noncentrosymmetric systems. Noncentrosymmetric materials with its unique properties are expected to host Majorana fermions, an emergent collective excitation of electrons \cite{topo1,topo2}.  An intrinsic antisymmetric spin-orbit coupling (ASOC) in these systems is expected to perturb the energy of electrons at the Fermi level and lead to nontrivial pairing of electrons \cite{rashba1,rashba2,rashba3}. Such a scenario is expected to host unconventional features like high upper critical field exceeding the Pauli limiting field, anisotropic/multiple superconducting gaps, time-reversal symmetry (TRS) breaking, magnetoelectric effects and topologically protected surface states \cite{Bauer2004,CePtSi,CeIS,LC,LNC1,MCE1,MCE2,Hsoc1,Hsoc3,topo3,topo4}. \\

One of the most intriguing yet bewildering properties of NCS superconductor is the appearance of a spontaneous field in the superconducting ground state and hence the broken time reversal symmetry. Among the NCS systems, the materials displayed a broken TRS include LaNiC$ _{2} $ \cite{LNC1}, La$ _{7} $X$ _{3} $ (X = Ir, Rh) \cite{LI,LR}, Re$ _{6} $X (X = Zr, Hf, Ti) \cite{RZ3,RH2,RT}, CaPtAs \cite{CaPtAs}. An interesting case study is of Re$ _{6} $X, where members from this family have shown spontaneous field, with almost the same strength, irrespective of the X element. This result hence undermines the proposed effects of ASOC on the appearance of unconventional nature. Moreover, the spontaneous field in the superconducting ground state of elemental Re powder with centrosymmetric structure has further added curiosity \cite{Re}. A very recent report on Re$ _{1-x} $Mo$ _{x} $ alloy has shown time reversal symmetry breaking for x = 0.12, which crystallizes in the centrosymmetric structure, while it showed a preserved time reversal state for all other compositions including the noncentrosymmetric $ \alpha $-Mn structure \cite{ReMo}. Also, Re$ _{3} $W \cite{RW} and Re$ _{3} $Ta \cite{RTa} both crystallizing in the $ \alpha $-Mn structure has also failed to show any spontaneous field in the superconducting state. Another binary compound, ReBe$ _{22} $ which crystallizes in centrosymmetric structure, has shown TRS is preserved \cite{RBe}. Although there are claims of the role played by elemental Re concentration, it is not yet verified. Hitherto, most of the Re-based systems microscopically studied has Re rich in concentration, except the ReBe$ _{22} $ and Re$ _{1-x} $Mo$ _{x} $ (x = 0.6) superconductors. Hence it is important to look for more Re-based superconductors with different Re concentrations, in particular, those with the noncentrosymmetric structure, to further understand the role played by structure and Re concentration.

In this study, we have investigated the nature of the superconducting ground state in NbReSi, containing an atomic ratio of 33 \% elemental Re. Superconductivity in this material was first reported in 1985 \cite{NReS}. NbReSi crystallizes in a noncentrosymmetric orthorhombic FeSiTi-type structure and enters the superconducting state below 6.29 K. This structure is a superstructure modification of hexagonal ZrNiAl-type structure. The hexagonal family has shown higher T$ _{c} $, with highest T$ _{c} $ = 13 K for ZrRuP, compared to the orthorhombic family. The high T$ _{c} $ in the hexagonal family is attributed to strong electron-phonon coupling strength, indicating the influence of structure on the superconducting properties of ternary equiatomic materials \cite{ZrRuP}. Here, we have used the muon spin rotation/relaxation technique ($ \mu $SR) to study the superconducting ground state of orthorhombic NbReSi. $ \mu $SR in the longitudinal geometry without any externally applied field is an excellent tool to detect any spontaneous magnetic field arising below the superconducting transition. While $ \mu $SR in transverse geometry with a small applied field can accurately determine the superconducting gap structure.
\begin{figure}
\includegraphics[width=1.0\columnwidth]{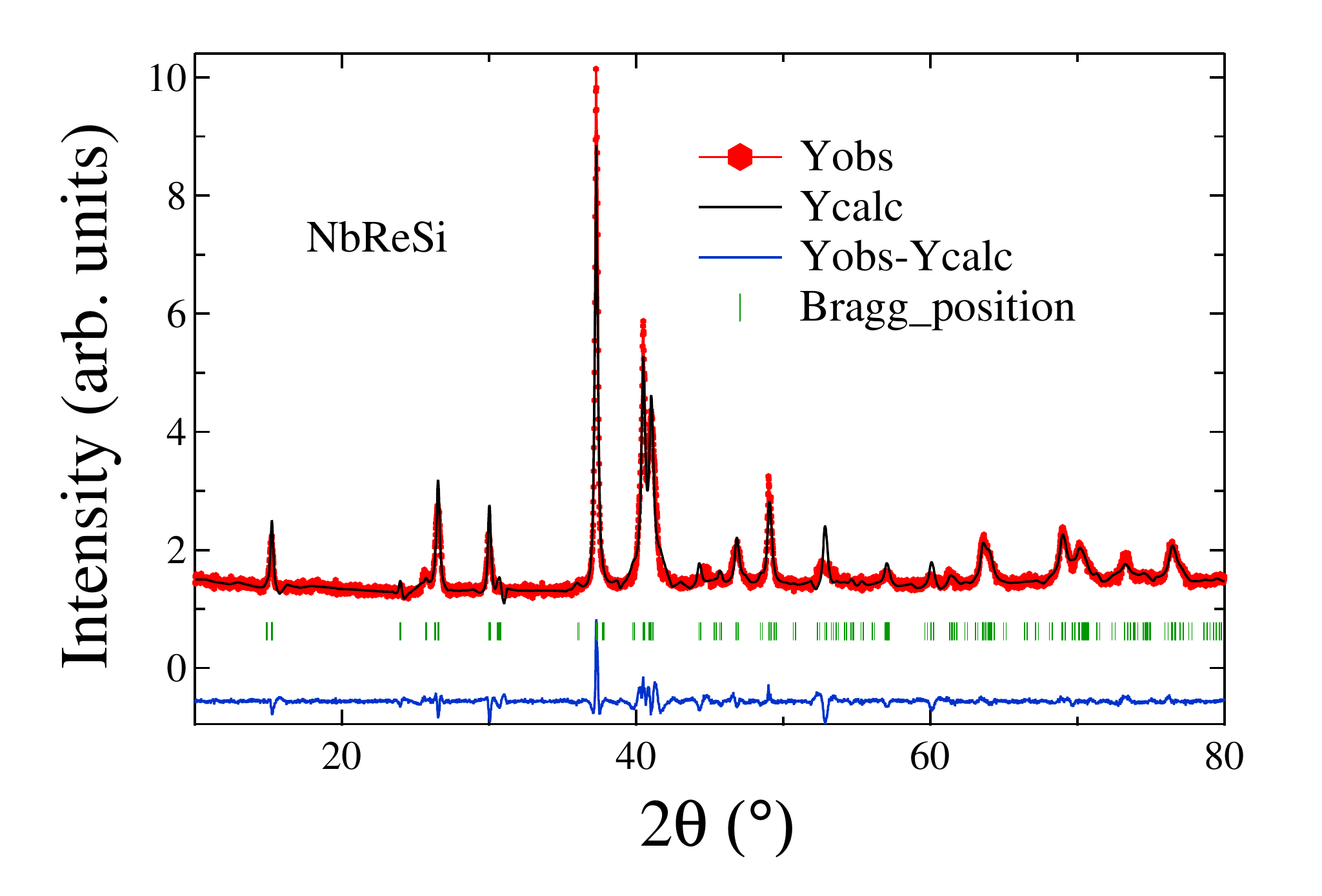}
\caption{\label{fig1}The powder X-ray diffraction pattern collected at ambient temperature and pressure. Solid black line is the Rietveld refinement to the data.}
\end{figure}
     
\section{EXPERIMENTAL METHODS}

A polycrystalline sample of NbReSi was prepared by arc melting stoichiometric amounts of the constituent elements on a water-cooled copper hearth under argon gas atmosphere. The samples were flipped and remelted several times with negligible mass loss to ensure the homogeneity of the ingot. The sample was wrapped in Ta foil, sealed in quartz ampules under vacuum, and annealed at 800 \textdegree{}C for one week. The sample characterization was done using the x-ray powder diffraction (XRD) on a PANalytical diffractometer using Cu K$_{\alpha}$ radiation ($\lambda$ = 1.54056 $\text{\AA}$). Magnetic susceptibility measurements were performed on a magnetic property measurement system (MPMS) superconducting quantum interference device (SQUID) magnetometer (Quantum Design). Magnetic measurements were taken in both zero field cooled (ZFC) mode and field cooled mode (FCC). The muon-spin relaxation/rotation ($\mu$SR) measurements were carried out using the MuSR spectrometer at the ISIS Neutron and Muon facility in STFC Rutherford Appleton Laboratory, United Kingdom. The powdered sample of NbReSi was mounted on a high-purity-silver plate using diluted GE varnish. The $\mu$SR measurements were performed in the longitudinal and transverse-field geometries. During measurement, spin-polarized muons were implanted into the sample. In the longitudinal configuration, the positrons were detected either in forward or backward positions along the axis of the muon beam. The asymmetry is calculated by
\begin{equation}
G_{Z}(t) = \frac{N_{\mathrm{B}}-\alpha N_{\mathrm{F}}}{N_{\mathrm{B}}+\alpha N_{\mathrm{F}}} 
\label{eqn1}
\end{equation}
\\
where $N_{\mathrm{F}}$ and $N_{\mathrm{B}}$ are the number of counts at the detectors in the forward and backward positions, and $\alpha$ is determined from calibration measurements taken with a small applied transverse magnetic field. In this configuration, measurements were made in zero field with the contribution from the stray fields at the sample position due to neighboring instruments, and the Earth's magnetic field is canceled to within $\sim$ 1.0 $\mu$T by using three sets of orthogonal coils. In the transverse configuration, a field was applied perpendicular to the direction of the muon beam, and the detectors were grouped into two orthogonal pairs.  A full description of the $\mu$SR technique may be found in ref. \cite{SLL}.
 \begin{figure}[t]
      	\includegraphics[width=1.0\columnwidth]{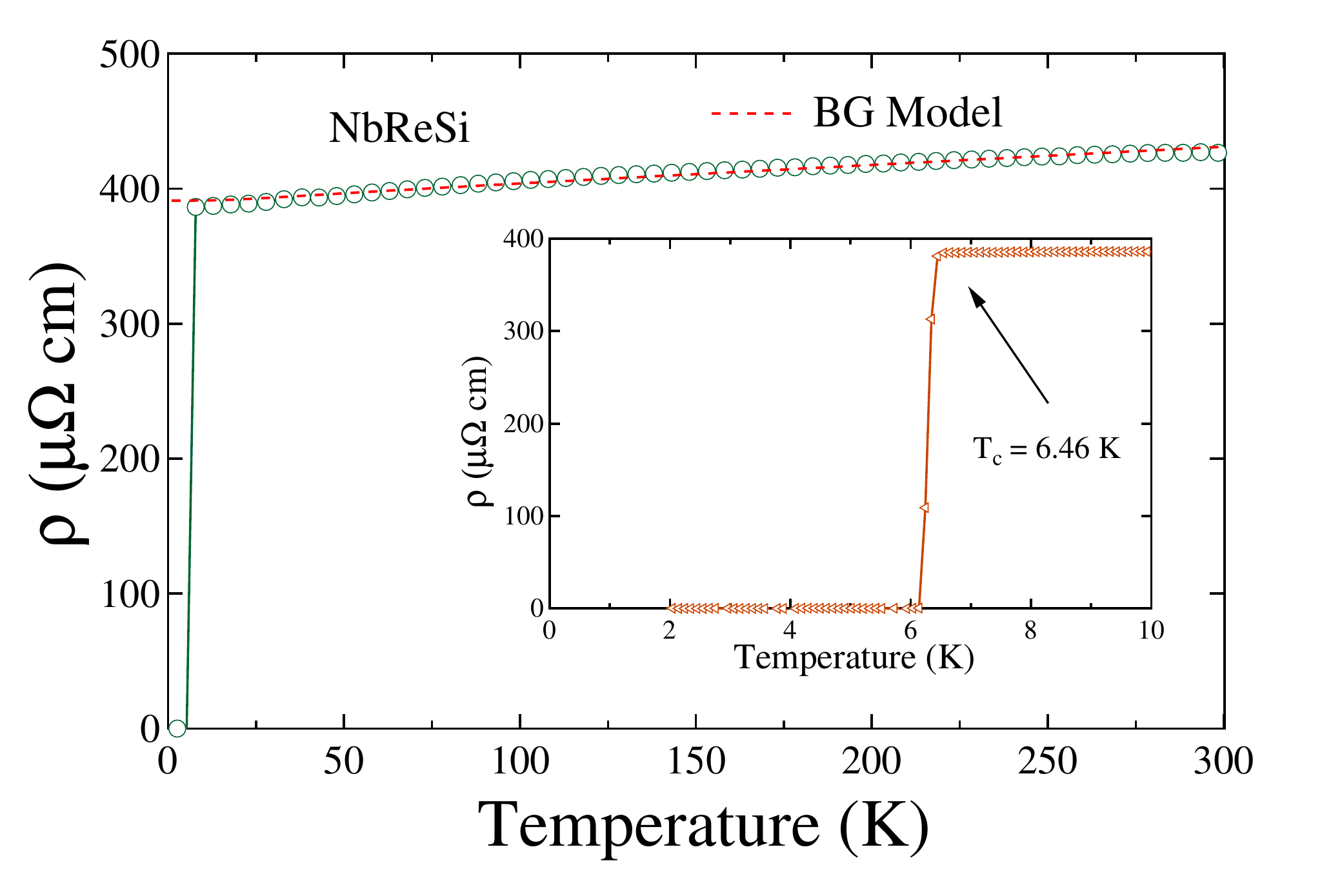}
      	\caption{\label{fig2}Resistivity data taken at zero field showing a metallic nature of NbReSi. The inset shows a sharp superconducting nature of sample below 6.46 K. The normal state resistivity is well described by the Bloch-Gruneisen (BG) Model and is shown by dotted red line. }
      \end{figure} 
\section{RESULTS AND DISCUSSION}
\begin{figure*}[t]
  	\includegraphics[width=2.0\columnwidth]{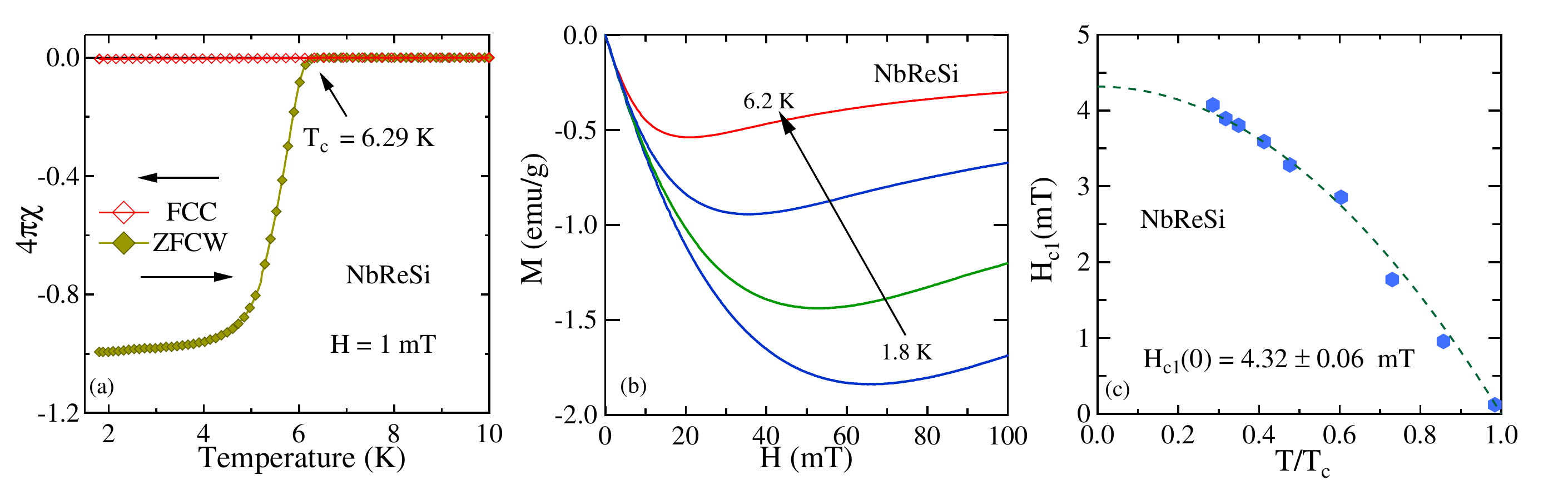}
  	\caption{\label{fig3}(a) DC magnetization data were taken at an applied field of 1 mT, showing a superconducting transition at 6.29(3) K. (b) Low field magnetization curves taken at different temperatures (c) Lower critical field, H$ _{c1} $ versus  normalized temperature for NbReSi. The dotted line showing an fit to the data using G-L equation gives H$ _{c1} $(0) = 4.32 $ \pm $ 0.06 mT.}
  \end{figure*} 
\subsection{Crystallography}
X-ray diffraction data collected at ambient pressure and temperature is shown in \figref{fig1}. Reitveld refinement of the data done using Fullprof \cite{Full} software shows no obvious impurity phase is present in sample. NbReSi adopts the orthorhombic FeSiTi-type structure which has the noncentrosymmetric space group, Ima2. The crystallographic parameters obtained are  a = 6.925(5) \text{\AA}, b = 11.671(2) \text{\AA}, c = 6.693(8) \text{\AA}, in good agreement with reported data \cite{NReS}.

 \subsection{Resistivity} 
 \figref{fig2} shows the temperature dependence of the resistivity, which for temperatures greater than 7K shows a small increase with temperature upto 300 K. However, below 10 K a sudden drop in resistivity around 6.46(7) K marks the onset of superconductivity in NbReSi. Zero resistivity was observed at 6.14 K which gives a transition width of $ \Delta $T = 0.32 K. The metallic nature of the sample can be inferred from the positive slope shown above T$ _{c} $. The resistivity data above superconducting transition can be well modeled by using the Bloch-Gruneisen model. This takes into account the resistivity arising due to electrons scattering from longitudinal acoustic phonon \cite{BG}. According to this model, the resistivity can be described as,

\begin{equation}
\rho(T) = \rho_{0} +  \rho_{BG}(T)
\label{BG1}
\end{equation} 

Where, $ \rho_{BG}(T) $ is defined as 

\begin{equation}
\rho_{BG}(T) = r\left(\frac{T}{\theta_{D}}\right)^{5}\int_{0}^{\Theta_{D}/T}\frac{x^{5}}{(e^{x}-1)(1-e^{-1})}dx
\label{para3}
\end{equation}

Here, $ \theta_{D} $ is the Debye temperature and r is a material dependent constant which depends on the plasma frequency and electron-phonon coupling strength. The results of the fit using this model yields: $ \rho_{0} $ = 391 $ \mu\Omega\;cm $, $ \theta_{D} $ = 99.88 K and $ r $ = 53.5 $ \mu\Omega\;cm $.

\subsection{Magnetization}
The bulk nature of superconductivity in NbReSi is tested by dc magnetization measurement in an applied field of 1 mT in  ZFC-FCC configuration. The onset of the superconducting state was observed at 6.29(3) K with a strong type-II nature indicated by flux pinning behavior below transition temperature in FCC measurement. The superconducting fraction was found to be close to 100\%, indicating a full superconducting fraction. \\

The field dependence of the dc magnetization was investigated at different temperatures between 1.8 K, and 6.2 K. The magnetization was seen to increase linearly with field up to a certain field value, and then deviates from linear dependence followed by the reverse in magnetization upon entering the vortex state. The point of deviation from linear behavior is taken as the value of the lower critical H$ _{c1}(T)$. The temperature variation of H$ _{c1} $ is well described by Ginzburg-Landau equation,  H$ _{c1} $(T) = H$ _{c1} $(0)(1-t$^{2}$), (t = T/T$ _{c} $), which gives H$ _{c1} $(0) = 4.32 $ \pm $ 0.06 mT.

\begin{figure}[t]
\includegraphics[width=1.0\columnwidth]{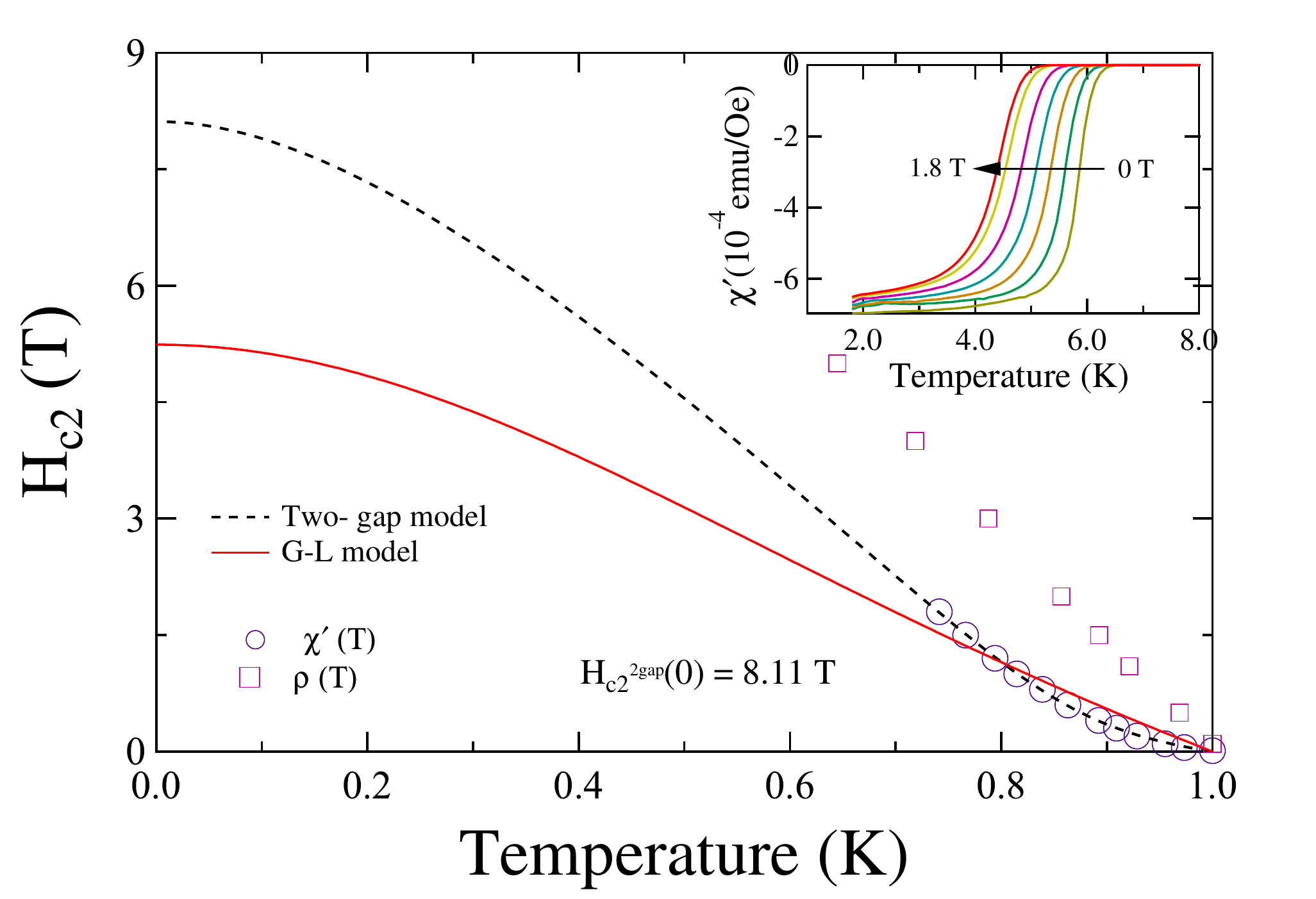}
\caption{\label{fig4} The temperature dependence of the upper critical field is estimated from ac susceptibility and resistivity measurements. The dotted line shows the fit to the data using the two-gap model, and the solid line shows the fit using the G-L equation. The estimated H$ _{c2} $(0) is 8.11 T.The inset shows the ac susceptibility data taken at different applied fields.  }
\end{figure}

The upper critical field for the sample was estimated using the resistivity and ac susceptibility data collected at a different applied field. The inset in Fig. \ref{fig4} shows the ac susceptibility data in different fields. The 50 \% of the drop in susceptibility is considered as the transition at the corresponding field. While, the uppercritical field estimated from resistivity has  shown comparatively higher value. This behavior can be attributed to surface or filamentary effects \cite{LPS,TReS}. The temperature dependence of the upper critical field estimated from ac susceptibility, as displayed in Fig. \ref{fig4} has shown an upward curvature at high field. This feature is reminiscent of a two-gap superconductor. Hence, a to fit the data using a two-gap model according to which the H$ _{c2} $(t) can be implicitly written in the parametric equation \cite{twogap},

\begin{multline}
ln\left(\frac{1}{t}\right) = \left[U(s) + U(\eta s) + \frac{\lambda_{0}}{w}\right] + \\  \left( \frac{1}{4} \left[U(s) - U(\eta s) - \frac{\lambda_{-}}{w}\right]^{2} +  \frac{\lambda _{eh}\lambda _{he}}{w^{2}} \right) ^{1/2}\\
H_{c2}   = \frac{ 2\phi_{0}Ts}{D_{e} } \;\;\;\;\;\;  \eta = \frac{D_{h}}{D_{e}}\\
U(s) = \psi(s+ 1/2) - \psi(1/2)\;\;\;\;\;\;\;\;\;\;\;\;\;\;
\label{whh}
\end{multline}

Here, $ \lambda_{-} $ = $ \lambda_{ee} - \lambda_{hh} $, $ \lambda_{0} = (\lambda_{-}^{2} + 4\lambda_{eh}\lambda_{he})$, $ w = \lambda_{ee}\lambda_{hh} - \lambda_{he}\lambda_{eh} $. The variables, $ \lambda_{ee}, \lambda_{hh}, \lambda_{eh}, \lambda_{he} $ are the matrix elements of the BCS coupling constants. $ D_{e} $ and $ D_{h} $ are the electron and hole diffusivity. $ \phi_{0} $ is the flux quantum and $ \psi (s) $ is the digamma function. An extrapolation using this model to T = 0 has yielded H$ _{c2} $(0) = 8.11 T. While, the G-L model failed to trace the data points underestimating the H$ _{c2} $(0) value. This value is greater than those reported for similar structured materials TaXSi (X = Re/Ru) \cite{TReS}. Nevertheless, a similar upward curvature in H$ _{c2} $(t) is expected for superconductors if nonmagnetic impurities or disorders are present \cite{twogap}. A high residual resistivity value and low value of residual resistivity ratio may point towards disorder in this system. Hence, to exactly elucidate the gap structure, measurements on high quality single crystal is required in future.    
\subsection{Muon spin rotation and relaxation measurements}
\begin{figure}
	\includegraphics[width=1.0\columnwidth]{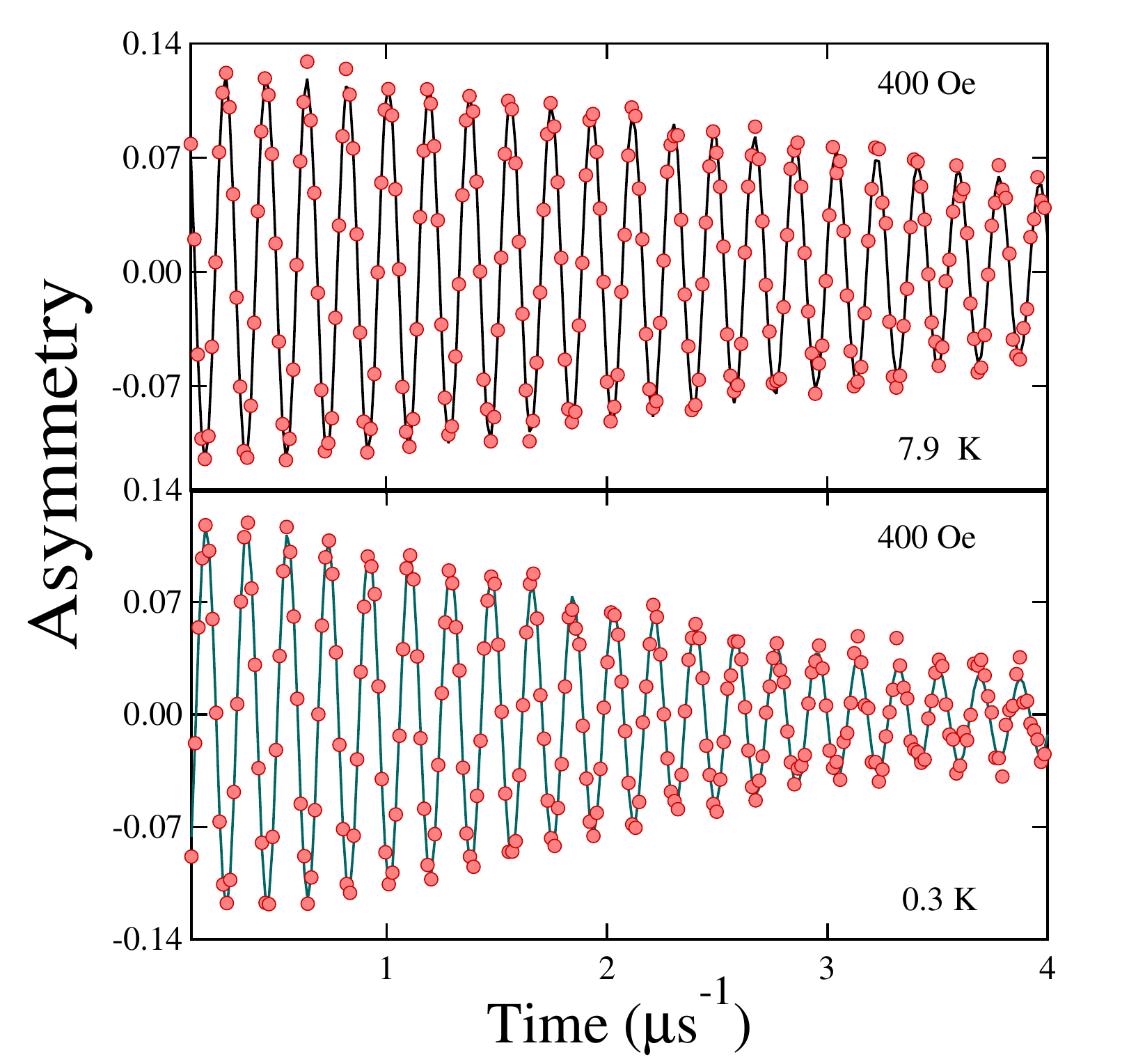}
	\caption{\label{fig5} Transverse field $ \mu $SR spectra collected at 40 mT at 7.9 K and 0.3 K. The increased depolarization of the muons in the sample due to the flux line lattice formation can be seen in (b) }
\end{figure}

To delve deeper, we have investigated the system using $ \mu $SR spectroscopy. TF-$ \mu $SR data was employed to investigate the superconducting gap structure of NbReSi. During this measurement, the magnetic field is applied orthogonal to the initial muon spin direction. For the formation of well ordered flux line lattice (FLL), the sample is cooled in an applied field of 40 mT, which is well above the lower critical field. The spectra below the transition temperature is seen rapidly depolarizing, which can be accounted by the inhomogeneous field distribution due to FLL (Fig. \ref{fig5}).    

The spectra can be well described by a combination of sinusoidally oscillating function damped with Gaussian relaxation and an oscillatory background term. 

\begin{eqnarray}
G_{\mathrm{TF}}(t) &=& A_{0}\mathrm{exp}\left(\frac{-\sigma^{2}t^{2}}{2}\right)\mathrm{cos}(\omega_{1}t+\phi)\nonumber\\&+&A_{1}\mathrm{cos}(\omega_{2}t+\phi) .
\label{eqn3}
\end{eqnarray} 

\begin{figure}
	\includegraphics[width=1.0\columnwidth]{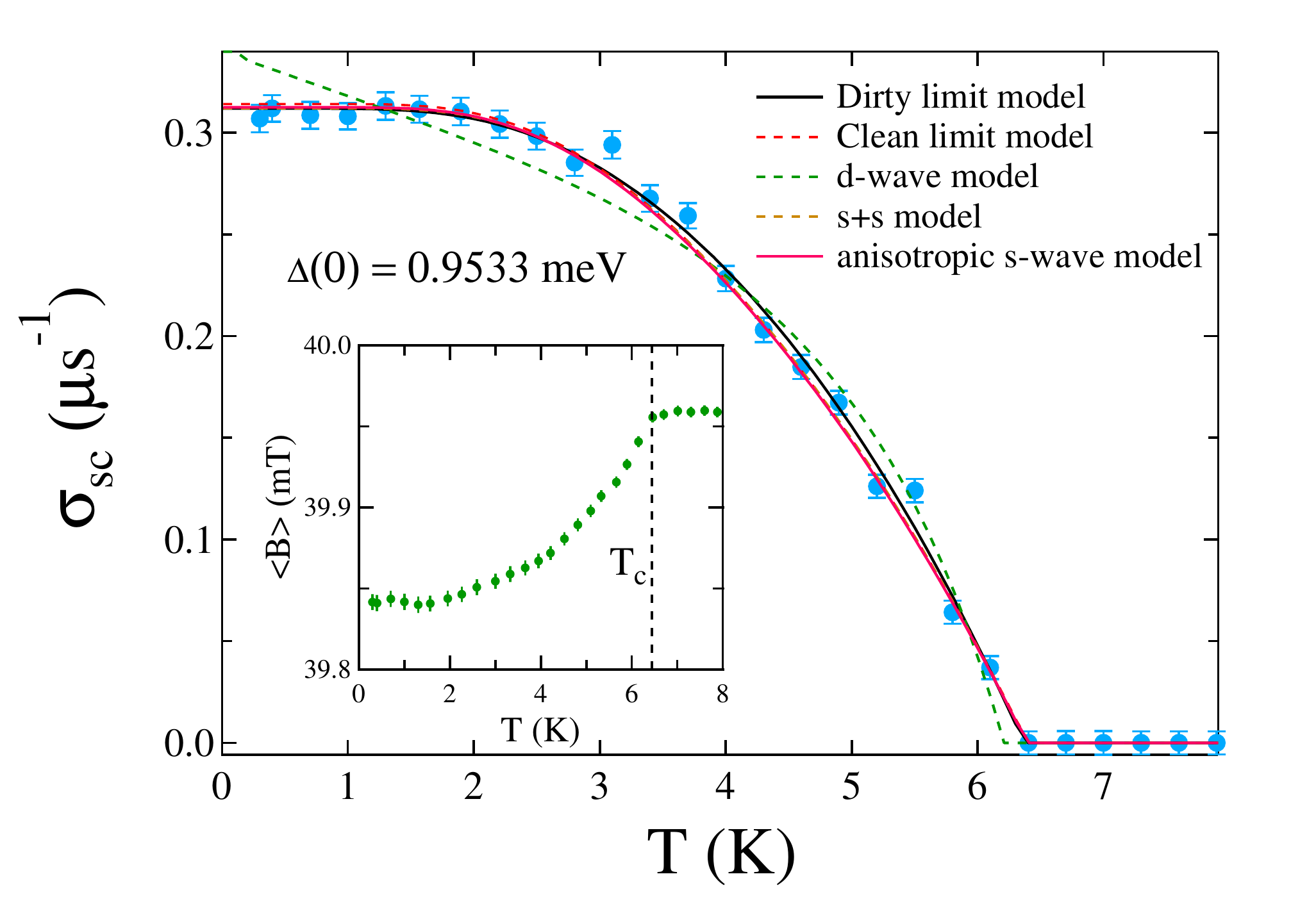}
	\caption{\label{fig6} Muon depolarization rate $ \sigma_{sc} $ collected at 400 Oe. The data collected at range of temperature across T$ _{c} $ is well traced by s-wave model giving the superconducting gap as $ \Delta $ (0) = 0.9533 meV. The inset shows the temperature dependence of the internal magnetic field. }
\end{figure}

Here the first term corresponds to the signal from sample, while the second corresponding to signal from the silver sample holder. A$ _{0} $ and A$ _{1} $ denotes the sample and background asymmetries, while $ \omega_{1} $ and $ \omega_{2} $ corresponds to the muon precession frequencies in the sample and background respectively. The depolarization rate, $ \sigma $ in \ref{eqn3} is comprised of two components, $ \sigma_{N} $ and $ \sigma_{sc} $. $ \sigma_{N} $ is accounted by the nuclear dipolar moments while $ \sigma_{sc} $ accounts for the depolarisation rate from the FLL. They are related by $\sigma^{2}$ = $\sigma_{\mathrm{sc}}^{2}+\sigma_{\mathrm{N}}^{2}$. The temperature dependence of $ \sigma_{sc} $ extracted using the above equation is plotted in \figref{fig6}. The depolarization has showed a plateau at low temperature after which it decreased upon increasing temperature and reaches zero at T$ _{c} $. This nature can be well followed by the s-wave model in the dirty limit as given by,

\begin{equation}
\frac{\sigma_{FLL}^{-2}( T )}{\sigma_{FLL}^{-2}(0)} = \frac{\Delta(T)}{\Delta(0)}\mathrm{tanh}\left[\frac{\Delta(T)}{2k_{B}T}\right] ,
\label{dirty s}
\end{equation}
where $\Delta(T)/\Delta(0) = \tanh\{1.82(1.018({T_{c}/T}-1))^{0.51}\}$ is the BCS approximation for the temperature dependence of the energy gap and $\Delta(0)$ is the gap magnitude at zero temperature. While in the clean limit, 
\begin{equation}
\frac{\sigma_{FLL}^{-2}(T)}{\sigma_{FLL}^{-2}(0)} = 1+2\int_{\Delta(T)}^{\infty}\left(\frac{\delta f}{\delta E}\right)\frac{E dE}{\sqrt{E^{2}-\Delta^{2}(T)}}  ,
\label{clean s}
\end{equation}
Here, $ f $ = [1+exp(E/k$_{B}T)]^{-1} $ is the Fermi function and $ \Delta (T) =  \Delta_{0}\delta (T/T _{c} )$ g($\phi$) . $\delta  (T/T _{c} )$ = $\tanh\{1.82(1.018({T_{c}/T}-1))^{0.51}\}$ is the temperature dependence of the energy gap. The term, $ g(\phi) $ accounts for the angular dependence of the gap function, where $ \phi $ is the azimuthal angle. $ g(\phi) $ can be substituted with (i) 1 for s-wave model, (ii) $\mid Cos(2\phi)\mid$ for d-wave model, and (iii) $(1+aCos(4\phi))/(1+a)$ for anisotropic gap, where $a$ represents the anisotropic parameter \cite{FeSe}. Among the different fitting models employed, the data was best described by dirty limit s-wave model ($ \chi^{2}_{norm.dirty} $ = 1.42 and $ \chi^{2}_{norm.clean} $ = 1.77 ). The low value of residual resistivity ratio also justifies the dirty limit nature of the sample. The dirty limit s-wave model fitting exactly retraces the path, giving the superconducting gap as $ \Delta(0) $ = 0.9533 meV. This gives the normalized superconducting gap as $ \Delta(0)/k_{B}T_{c} $ = 1.726, showing the moderately coupled nature of the sample. While, in the anisotropic model, the anisotropic parameter, a, has converged to a very small value 0.008, which rules out any anisotropic nature of the gap. While in the d-wave model $ \chi^{2} $ = 7.4, negates any d-wave nature. In order to check any multi-gap nature as point out by susceptibility measurements,  We have also performed a two-gap model fitting, where the total depolarization is expressed a sum of two components,

\begin{equation}
\frac{\sigma_{FLL}^{-2}(T)}{\sigma_{FLL}^{-2}(0)} = \omega \frac{\sigma_{FLL}^{-2}(T, \Delta_{0,1})}{\sigma_{FLL}^{-2}(0,\Delta_{0,1})} + (1 - \omega) \frac{\sigma_{FLL}^{-2}(T, \Delta_{0,2})}{\sigma_{FLL}^{-2}(0,\Delta_{0,2})}
\label{clean s}
\end{equation}

Here, $ \Delta_{0,1} $ and $ \Delta_{0,2} $ are the zero temperature values of the two gaps. Using this model gives $ \omega $ = 1, which converges into single band model. Hence, in conclusion, the TF-$ \mu $SR data shows the isotropic s-wave nature of NbReSi. However, it is to be noted that the measurement is performed at an applied field of 400 Oe, which could suppress a two-gap nature. Hence further detailed measurements on a single crystal sample are necessary at low applied fields.   

\begin{figure}
	\includegraphics[width=1.0\columnwidth]{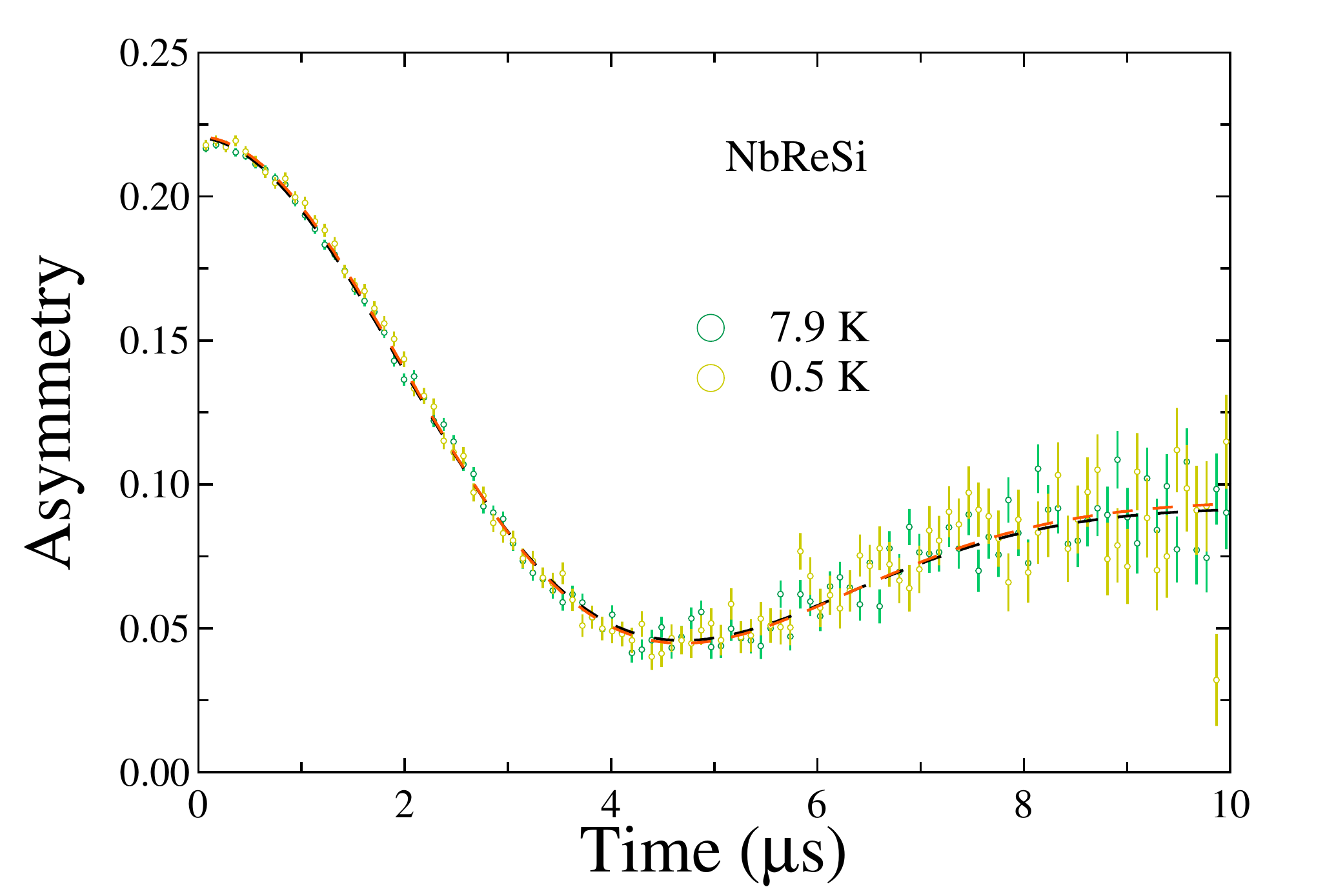}
	\caption{\label{fig7} Longitudinal asymmetry spectra collected at zero field at two different temperatures above and below T$ _{c} $. Both spectra was seen following the same path indicating the absence of any spontaneous field at superconducting phase. The dotted lines are the fit to the data using the Eq. \ref{eqn2}. }
\end{figure}

Muons being very sensitive to small magnetic fields can be employed to investigate the system further. A measurement performed in the zero field  and longitudinal geometry (ZF $ \mu $SR) can detect spontaneous magnetic field, if present, below T$ _{c} $. ZF $ \mu $SR spectra collected at two temperature above and below T$ _{c} $ is shown in \figref{fig7}. Absence of any processional signal rules out the presence of any coherent long-range magnetic ordering.  In the absence of any atomic moments, the muon depolarization is solely due to randomly oriented nuclear moments, which can be best described by a function,

\begin{equation}
G(t)= A_{1}\mathrm{exp}(-\Lambda t)G_{\mathrm{KT}}(t)+A_{\mathrm{BG}} ,
\label{eqn2}
\end{equation}

where A$ _{1} $, A$ _{BG} $ represent the sample and background asymmetry respectivly, while $ \Lambda $ accounts for the electronic relaxation rate. $ G_{KT} $ is the standard Kubo-Toyabe function given by \cite{KT1},

\begin{eqnarray}
G_{\mathrm{KT}}(t) &=&\frac{1}{3}+\frac{2}{3}(1-\sigma^{2}_{\mathrm{ZF}}t^{2})\mathrm{exp}\left(\frac{-\sigma^{2}_{\mathrm{ZF}}t^{2}}{2}\right),
\label{eqn6}
\end{eqnarray}

$ \sigma_{ZF} $ accounts for the relaxation due to static, randomly oriented local fields associated with nuclear moments at the muon site. An identical relaxation behavior is shown by the spectra collected at temperatures above and below the T$ _{c} $. This is evidence of the absence of spontaneous field originating in the superconducting state and hence a time reversal symmetry is preserved in the superconducting ground state.

\section{CONCLUSION}
In this particular study, we have characterized the ternary noncentrosymmetric compound NbReSi by x-ray diffraction, magnetization, resistivity, and muon spectroscopy measurements. The results indicate NbReSi enters the superconducting state below T$ _{c} $ = 6.4 K and shows a strong type-II behavior. The upper critical field curve estimated from ac susceptibility data has shown an upward curvature, reminiscent of two-gap superconductivity. However, this nature can also arise due to disorder/inhomogeneity in the system. The transverse field $ \mu $SR measurements showed a single BCS type isotropic superconducting gap with $ \Delta(0)/k_{B}T_{c} $ = 1.726, showing a moderately coupled nature. Thus indicating that disorder/inhomogeneity may give rise to the upward curvature of H$_{c2}$. Hence further detailed measurements on single crystal samples are required to conclusively state the gap structure in NbReSi. Furthermore, the zero field measurements showed time reversal symmetry preserved superconducting ground state. This result asks for further studies in Re-based compounds to understand the skeptical role of elemental Re in many unconventional superconducting systems.

\section{Acknowledgments} R.~P.~S.\ acknowledges Science and Engineering Research Board, Government of India for the Core Research Grant CRG/2019/001028. We thank Newton Bhabha for funding
and ISIS, STFC, UK, for the muon beam time to conduct the
μSR experiments.

\end{document}